\documentclass[pre,twocolumn,groupedaddress,showpacs,showkeys,amsmath,amssymb,floatfix]{revtex4-1}

\usepackage{graphicx}
\usepackage{dcolumn}
\usepackage{bm}

\begin{document}

\title{Percolation threshold on planar Euclidean Gabriel Graphs}
\author{C. Norrenbrock}
\email{christoph.norrenbrock@uni-oldenburg.de}
\affiliation{Institut f\"ur Physik, Universit\"at Oldenburg, 26111 Oldenburg, Germany}
\date{\today}

\begin{abstract}
In the present article, numerical simulations have been performed to find the bond and site percolation thresholds on two-dimensional Gabriel graphs (GG) for Poisson point processes. 
GGs belong to the family of ``proximity graphs'' and are discussed, e.g., in context of the construction of backbones for wireless ad-hoc networks.
In order to find the critical points, finite-size scaling analyses have been performed for several observables.
The critical exponents obtained this way verify that the associated universality class is that of standard $2D$ percolation.
\end{abstract}

\maketitle

\section{Introduction}
\label{sect:intro}

\begin{figure}[t!]
    \centerline{\includegraphics[width=1.0\linewidth]{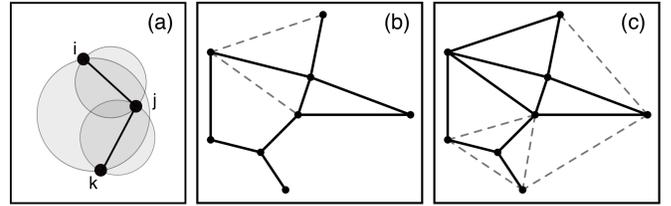}}
    \caption{(a) sketch of the construction rule for designing a GG. 
    Node-pair $i$,$j$ as well as $j$,$k$ are linked, since ${\rm circ}(i,j)$ and ${\rm circ}(j,k)$ do not contain further nodes. 
    In contrast, ${\rm circ}(i,k)$ contains node $j$, thus $i$ and $k$ are not connected directly by an edge. 
    (b) GG of size $N=8$. If the dashed edges got removed, the graph would change to a relative-neighborhood graph. 
    (c) Delaunay triangulation for the same node-set as in (b). 
    Removing the dashed edges would result in a GG.
    \label{fig:gg_final}}
\end{figure}

\begin{figure}[t!]
    \centerline{\includegraphics[width=1.0\linewidth]{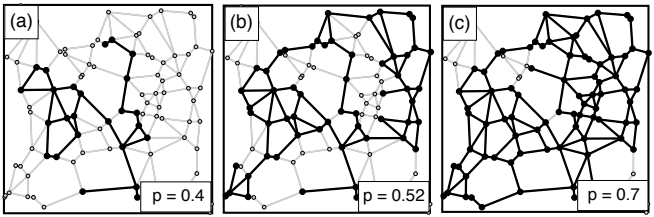}}
    \caption{Illustration of bond percolation on a GG of size $N=100$. 
    The fraction of occupied edges is given by $p$. The largest cluster is highlighted black. 
    All other edges, even those which are occupied, but do not belong to the largest cluster, are colored gray. 
    (a) $p=0.4$ (subcritical), (b) $p=0.52$ (slightly above the critical point $p_c=0.5167(8)$) and $p=0.7$ (supercritical)
        \label{fig:percol_final}}
\end{figure}

In this article standard percolation on Gabriel graphs \cite{Gabriel1969} is under scrutiny.
Standard percolation addresses the question of connectivity \cite{Stauffer1979,Stauffer1992}. 
E.g.\ in respect to site percolation, each site on a lattice is occupied randomly with probability $p$ or empty with probability $1-p$. 
Then, the pivotal objects of interest are clusters composed of occupied and adjacent sites.
The geometrical properties of these clusters change by shifting $p$ from small values to values close to $1$.
If $p$ is below a certain value $p_c$, the clusters will be small and disconnected. For $p>p_c$ instead, there will be basically one large cluster which covers almost the whole lattice.
Due to its fundamental nature and its adaptability to many different systems, there is already an abundance of literature about percolation.
E.g., it has been studied in context of marketing \cite{Goldenberg2000}, forest fire \cite{Henley1993} or disease spreading \cite{Moore2000}.
But also percolation by itself has been investigated extensively \cite{Newman2000,Stauffer1979,Stauffer1992}.
Besides simple configurational statistics there are many variants of the percolation problem such as the negative-weight percolation problem \cite{Melchert2008} or domain-wall excitations in $2D$ spin glasses \cite{Cieplak1994,Melchert2007} that requires a high degree of optimization.
Furthermore, more related to this work, the percolation phenomena has also been studied extensively on planar random graphs and their respective duals \cite{Hsu1999,Becker2009,Kownacki2008}.
E.g.\ in \cite{Becker2009}, the $2D$ Voronoi graph and its dual the Delaunay triangulation, which is a super graph of the Gabriel graph, are considered.\\
In this article we study standard percolation, i.e.\ bond and site percolation, on the Gabriel graph (GG) for a set of $N$ randomly placed points in the planar Euclidean space.
In general, a graph $G=(V,E)$ consists of a node set $V$ and an edge set $E$ \cite{Essam1970}.
In respect to a node set in a two-dimensional plane there will only exist an undirected edge $e_{i,j}\in E$ between two different nodes $i,j\in V$ in the Gabriel graph if the following condition is fulfilled: 
\begin{equation}
    d^2_{i,j} < d^2_{i,k} + d^2_{j,k}\hspace{0.5cm}\forall\;k\in V \backslash \{i,j\},
    \label{eq:gabdef}
\end{equation}
where $d_{i,j}$ denotes the Euclidean distance between $i$ and $j$ \cite{Santi2005}.
This means there will be an edge between two nodes $i$ and $j$ only if the disk which has the connecting line between $i$ and $j$ as its diameter contains no further nodes $k\in V \backslash \{i,j\}$. 
Furthermore, nodes in the Gabriel graph are never linked to themselves. 
The described linking rule is sketched in Fig.\ \ref{fig:gg_final}(a).
The gray-shaded areas between two nodes illustrate the aforementioned disc used for checking whether the nodes get connected or not. 
As shown in the figure both node-pairs $i$,$j$ and $j$,$k$ are linked since their respective disc is empty. 
In contrast, the nodes $i$ and $k$ are not connected, because their respective disc embeds node $j$. 
Consequently, regarding this small example, the GG is: $G=(\{i,j,k\},\{(i,j),(j,k)\})$.\\
Due to its construction the Gabriel graphs exhibit some characteristic properties, e.g.\ rather short paths between all node-pairs, that are of interest in the context of communication networks. 
For that reason, GGs are discussed of being potential candidates for ``virtual backbones'' in ad-hoc networks, i.e.\ collections of radio devices without fixed underlying infrastructure \cite{Karp2000,Santi2005,Bose2001,Kuhn2003}.
GGs have also been applied a lot in geographic variation studies in biology \cite{Sokal1980,Sokal1978,Selander1975}.
In Ref.\ \cite{Melchert2013-2} the bond and site percolation thresholds of an Euclidean relative neighborhood graph (RNG) for a planar point set are studied.
In that article it is emphasized that the RNG is designed as a supergraph of the minimum weight spanning tree (MST) \cite{Michailidis2005}, i.e.\ regarding the same point set, all edges contained in the MST are also included in the RNG, and a subgraph of the Delaunay triangulation (DT) \cite{Delaunay1934}. 
The containment principle due to Fisher \cite{Fisher1961} states that if $G^\prime$ is a subgraph of $G$, the bond and site percolation thresholds exhibit $p_c^{G^\prime}\geq p_c^G$. 
It is shown in Ref.\ \cite{Melchert2013-2} that this is true for the ${\rm MST} \subset {\rm RNG} \subset {\rm DT}$ hierarchy.
The linking rule of the GG implies that the GG is a supergraph of the RNG and a subgraph of the DT, which is illustrated in Fig.\ \ref{fig:gg_final}(b) and Fig.\ \ref{fig:gg_final}(c) for a node set composed of $N=8$ nodes.
Consequently, it is expected that $p^{\,{\rm RNG}}_{c,{\rm bond}} = 0.771(2)\;\text{(Ref.\ \cite{Melchert2013-2})} \geq p^{\,{\rm GG}}_{c,{\rm bond}} \geq p^{\,{\rm DT}}_{c,{\rm bond}} = 0.333069(2)\;\text{(Ref.\ \cite{Becker2009})}$ as well as $p^{\,{\rm RNG}}_{c,{\rm site}} = 0.796(2)\;\text{(Ref.\ \cite{Melchert2013-2})} \geq p^{\,{\rm GG}}_{c,{\rm site}} \geq p^{\,{\rm DT}}_{c,{\rm site}} = 0.5\;\text{(Ref.\ \cite{Bollobas2006})}$. 
This will be confirmed in Sect.\ \ref{sect:results}.
A straightforward implementation to design the GG for a given set of points would terminate in time $\mathcal{O}(N^3)$, since for each node-pair all other nodes must be taken into account to check Eq.\ \ref{eq:gabdef}.
A substantially faster algorithm is introduced in Ref.\ \cite{Toussaint1980} to design the RNG, which, however, also works to construct the GG.
This algorithm utilizes that the GG is a subgraph of the DT. For points on two-dimensional surfaces, the DT can be computed fast terminating in time $\mathcal{O}(N\log(N))$ \cite{Preparata1985,qhull}. After computing the DT, Eq.\ \ref{eq:gabdef} can be checked for each of its edges resulting in a worst case running time of $\mathcal{O}(N^2)$.
However, the running time can further increased by implementing ``range queries'' (see Sect.\ \ref{sect:gabriel}) leading to $\mathcal{O}(N\log(N))$.
Note that the fast implementation works solely in the planar case and for the Euclidean metric. 
In principle, the GG as well as the DT or RNG can be designed for other metrics and in other dimensions. 
We study bond and site percolation here. 
Thus, for a given instance of the GG a fraction $p$ of the edges (bond percolation) or nodes (site percolation) gets occupied. 
Then we consider the geometrical properties of the appearing clusters consisting of adjacent nodes that are either connected by occupied edges (bond percolation) or which are occupied by themselves (site percolation).
For three different values of $p$ the largest cluster is illustrated in Fig.\ \ref{fig:percol_final} regarding bond percolation for an instance of $N=100$ nodes.
Since it is known that there are nontrivial percolation thresholds for Poisson point processes on the DT \cite{Becker2009} and that the percolation thresholds of the MST are equal to $1$, the question what subgraphs of the DT exhibit nontrivial percolation thresholds has been addressed in Ref.\ \cite{Billiot2010}.
Regarding the GG, this has been answered by Bertin et al.\ in Ref.\ \cite{Bertin2002}.
In the latter article it is proven analytically that there are nontrivial percolation thresholds for Poisson point processes. 
Also numerical simulations have been made to obtain rough estimates for the critical points ($p_{c,{\rm site}}=0.52$ and $p_{c,{\rm bond}}=0.64$).
Here, we measure the critical points numerically with much higher accuracy using finite-size scaling analyses.
For consistency, we also measure several critical exponents which are expected to be equal to those of standard $2D$ percolation due to universality.

The remainder of this article is organized as follows.
In Sect.\ \ref{sect:gabriel}, the algorithm for designing Gabriel graphs on randomly placed nodes is introduced.
In Sect.\ \ref{sect:results}, we present our numerical results.
Finally, we conclude with a summary in Sect.\ \ref{sect:summary}.

\section{Construction of Gabriel graphs}
\label{sect:gabriel}

In a naive implementation of the GG each pair of nodes $i$,$j$ must be considered successively terminating a a running time $\mathcal{O}(N^2)$.
Additionally, for each node-pair all other nodes $k\in V\backslash \{i,j\}$ have to be taken into account to check Eq.\ \ref{eq:gabdef} resulting in a running time of $\mathcal{O}(N^3)$.
However, realizing that one single node lying on the disc between $i$ and $j$ is sufficient to reject edge $(i,j)$, one can obtain a substantial speed-up.
Here, we consider points on a two-dimensional surface and we apply the Euclidean metric to determine their distances.
For this case, we can design GGs by means of an efficient algorithm \cite{Toussaint1980} which utilizes that the GG is a subgraph of the DT.
For a given point set, the DT can be constructed first. 
We did this using the Qhull computational geometry library \cite{qhull} terminating in time $\mathcal{O}(N\log(N))$.
Subsequently, each edge of the DT can be checked and possibly deleted by examining the disc between the adjoined nodes. 
For this we used the ``cell-list'' method \cite{Melchert2013-2}.
After placing $N$ nodes randomly on a $[0,1]\times[0,1]$ surface, we divide the unit square into $L\times L$ cells ($L=\sqrt{N}$) and equip each cell with a list of all nodes contained in.
Consequently, if the disc between two nodes is to be checked for being blank, just the nodes in the respective cells must be taken into account.
A running-time analysis and a more precise description of the cell-list method is given in Ref.\ \cite{Melchert2013-2} in the context of RNGs.

\section{Results}
\label{sect:results}

The current section is divided into two subsections.
First, in Sect.\ \ref{sect:bond}, the general concepts of finite-size scaling analyses will be explained and different observables to study the percolation phenomenon will be introduced. The analysis regarding bond percolation will be illustrated in detail. 
The results of site percolation are presented in Sect. \ref{sect:site} in a more brief manner, since the analysis is conceptually equal to that of bond percolation.

The Euclidean GGs are constructed using the efficient DT-based algorithm (c.f.\ Sect.\ \ref{sect:gabriel}) for planar sets of $N=12^2\dots192^2$ points which are placed randomly on a unit square. All results are averaged over $2000$ independent instances of the GG indicated by $\langle\dots\rangle$.

\subsection{Results for bond percolation on planar GGs}
\label{sect:bond}

For a given GG instance, the bond percolation (as well as site percolation) problem has been simulated using the highly efficient, union-find based algorithm by Newman and Ziff \cite{Newman2000,Newman2001}.
In the vicinity of the expected values of the percolation thresholds ($p\in[p^{{\rm expect}}_c-0.05,p^{{\rm expect}}_c+0.05]$) several observables $y(p,L)$ have been monitored.
Following a common scaling assumption \cite{Stauffer1992}, these observables can be rescaled according to
\begin{equation}
    y(p,L)=L^{-b}f[(p-p_c)L^{1/\nu}],
    \label{eq:scalassump}
\end{equation}
where $p_c$ denotes the critical point, and $\nu$ and $b$ represent dimensionless critical exponents.
$f[\cdot]$ is an unknown scaling function.
It becomes evident from Eq.\ \ref{eq:scalassump} that all data points of $L^by(p,L)$ have to lie on one single curve if $p_c$, $\nu$ and $b$ are chosen correctly.
Thus, in order to find the critical point, one just has to measure $y(p,L)$ for different values of $p$ and $L$. 
Then plotting $L^by(p,L)$ against $\epsilon\equiv(p-p_c)L^{1/\nu}$, and adjusting the unknown constants sufficiently will result in a ``data collapse'' indicating that the right values of the constants are found.
In this way the finite-size effects are exploited in order to estimate the critical parameters.
However, note that the scaling behavior of small systems might differ slightly from Eq.\ \ref{eq:scalassump} \cite{Binder2002}. 
Therefore, it might become necessary for some observables to ignore small systems when collapsing the data.
All data collapses in this article have been made by means of a computer-assisted scaling analysis \cite{Melchert2009-2}.
The results obtained for bond and site percolation are listed in Tab.\ \ref{tab:critp}.
\begin{table}[b]
    \begin{tabular}{lllll}\hline\hline\\[-9pt]
    Type & $p_c$ & $\nu$ & $\beta$ & $\gamma$\\[2pt]\hline\\[-9pt]
    GG-BP\hspace{14pt} & 0.5167(6)\hspace{14pt}& 1.33(6)\hspace{14pt} & 0.139(9)\hspace{14pt} & 2.39(5)\\[2pt]
    GG-SP\hspace{14pt} & 0.6348(8)\hspace{14pt} & 1.33(13)\hspace{14pt} & 0.139(8)\hspace{14pt} & 2.42(5) \\[2pt]\hline\hline
    \end{tabular}
    \caption{
        Critical points and critical exponents for bond percolation (GG-BP) and site percolation (GG-SP) on the Euclidean GG for planar point-sets. 
        The listed estimates for $p_c$, $\nu$ and $\beta$ are obtained by the scaling analyses of the order parameter (c.f.\ Fig.\ \ref{fig:bond_all} (b) regarding GG-BP). 
        The listed estimates for $\gamma$ are obtained by considering the averaged size of the finite clusters (c.f.\ Fig.\ \ref{fig:bond_all} (d) regarding GG-BP). 
        \label{tab:critp}}
\end{table}

\begin{figure}[t!]
    \centerline{\includegraphics[width=1.0\linewidth]{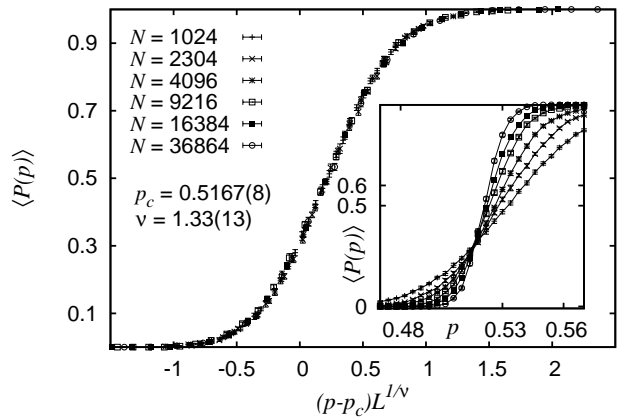}}
    \caption{Bond-percolation probability $P(p)$ over $p$ in the vicinity of the critical point on GG for planar point sets (inset). Using the scaling assumption (Eq. \ref{eq:scalassump}), the data are collapsed to one curve providing estimates of $p_c$ and $\nu$ (main plot). The data for each system size have been obtained by averaging over $2000$ realizations of the disorder, i.e.\ different point sets.
    \label{fig:bond_percol}}
\end{figure}

\textit{Percolation probability.} 
In order to measure the percolation probability we first determine the $L$ nodes that are closest to the left border of the unit square. 
We do the same for the right, top and bottom border.
These nodes are considered as being the border of the planar graph.
Then a cluster consisting of adjacent nodes linked by occupied edges will be considered as percolating if it contains at least one node from each of these four borders.
E.g.\ the clusters depicted in Fig.\ \ref{fig:percol_final}(b) and Fig.\ \ref{fig:percol_final}(c) percolate.
The finite size scaling analysis regarding the percolation probability $P(p)$ is depicted in Fig.\ \ref{fig:bond_percol}.
Since $P(p)$ is a dimensionless quantity, $b=0$ is set in Eq.\ \ref{eq:scalassump}.
The scaling assumption just applies in the vicinity of the critical point. Thus, not all data should be taken into account for finding the data collapse.
Regarding $P(p)$ the region $\epsilon\in[-0.25,0.5]$, representing the ``critical scaling window'', has been considered only. 
The estimates $p_c=0.5167(8)$ and $\nu=1.33(13)$ provide the best data collapse with quality $S=0.69$. The quality $S$ denotes the mean-square distance of the data points to the unknown scaling function in units of the standard error \cite{Melchert2009-2}.
$\nu$ describes the correlation length exponent and matches well with the known value of standard $2D$ percolation: $\nu=4/3\approx1.333$.

\begin{figure*}[t!]
    \centerline{\includegraphics[width=0.97\linewidth]{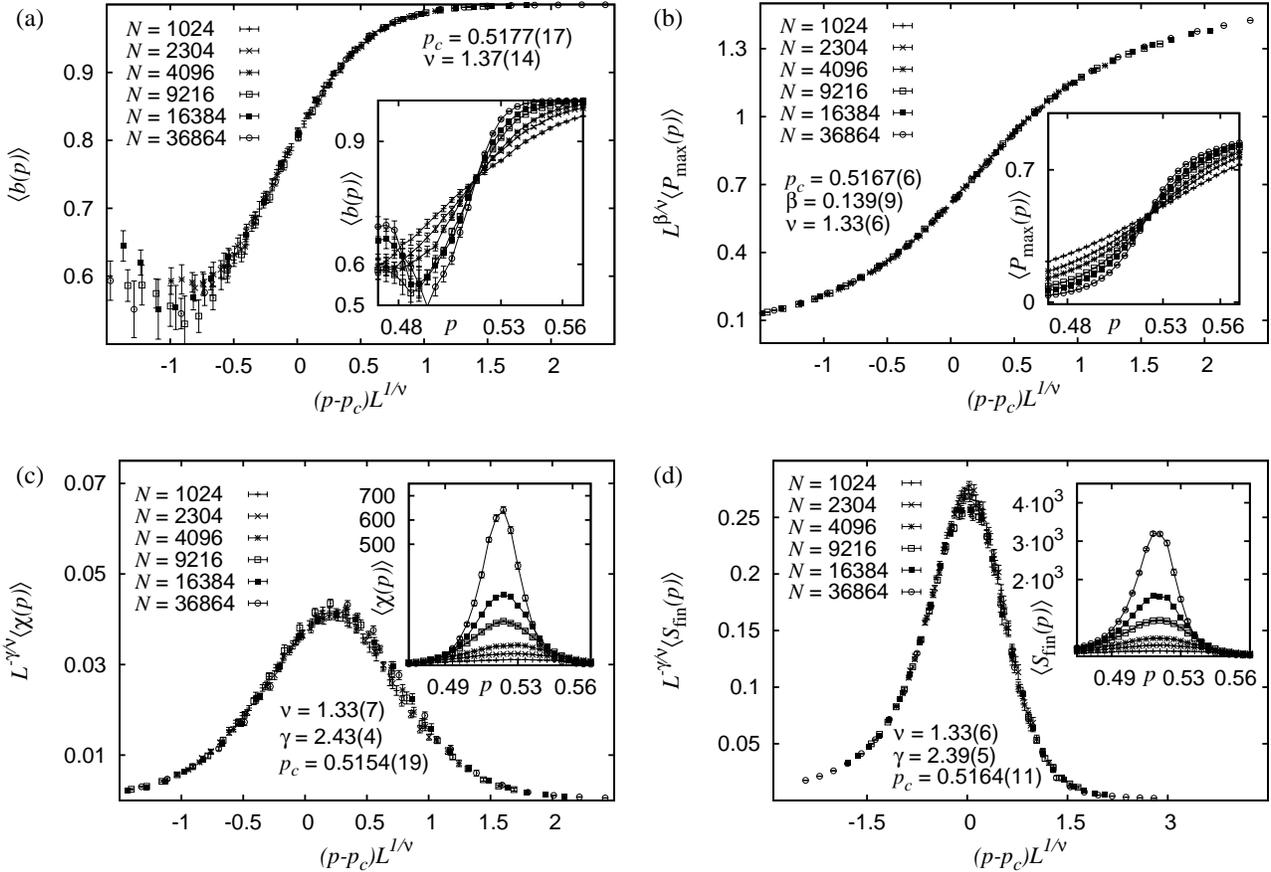}}
    \caption{Finite-size scaling analyses for different observables obtained by considering the bond-percolation problem on GGs for planar sets of up to $N=36864$ points.
    All data have been obtained by averaging over $2000$ instances. 
    The insets illustrate the raw data. The main plots show the collapsed data obtained following a rescaling according to Eq.\ \ref{eq:scalassump}. 
    (a) Binder ratio (c.f.\ Eq.\ \ref{eq:binder}), 
    (b) order parameter, i.e.\ the relative size of the largest cluster, 
    (c) fluctuations of the order parameter (c.f.\ Eq.\ \ref{eq:suscept}), 
    and (d) averaged size of the finite clusters (c.f.\ Eq.\ \ref{eq:sfin}).
    For obtaining the estimates $p_c$, $\nu$, $\beta$ and $\gamma$, all systems sizes $N=1024\ldots36864$ have been considered. 
    However, this is not true for (d), where just $N=9216\ldots36864$ have been taken into account.
    \label{fig:bond_all}}
\end{figure*}

\textit{Order parameter statistics.}
We also monitor the relative size of the largest cluster $s_{\rm max}$, i.e.\ the number of nodes in the largest cluster divided by $N$.
A common quantity of interest in this regard is the dimensionless \textit{Binder ratio}
\begin{equation}
    b(p)=\frac{1}{2}\bigg[3-\frac{\langle s^4_{\rm max}(p)\rangle}{\langle s_{\rm max}^2(p)\rangle^2}\bigg],
    \label{eq:binder}
\end{equation}
which features a nice crossing point for different $N$ at the critical point. This can be seen in the inset of Fig.\ \ref{fig:bond_all}(a).
Considering $\epsilon\in[-0.75,1]$, we find $p_c=5177(17)$ and $\nu=1.37(14)$ with quality $S=0.76$.
Again, the estimate of $\nu$ is in accordance with the known literature value.
Next we consider the \textit{order parameter}
\begin{equation}
    P_{\rm max}(p) = \langle s_{\rm max}(p)\rangle.
    \label{eq:orderp}
\end{equation}
The best data collapse yields $p_c=0.5167(6)$, $\nu=1.33(6)$ and $\beta=0.139(9)$ with quality $S=0.59$ by considering $\epsilon\in[-0.25,0.75]$.
The estimates are in good agreement with the analytical values $\nu$ and $\beta=5/36\approx0.139$. 
We also consider the \textit{order parameter fluctuations}
\begin{equation}
    \chi(p) = N\big[\langle s^2_{\rm max}(p)\rangle-\langle s_{\rm max}(p)\rangle^2\big].
    \label{eq:suscept}
\end{equation}
This quantity provides a further critical exponent whose value is also known: $\gamma=43/18\approx2.389$.
The best data collapse for $\epsilon\in[-1,0.25]$ provides $p_c=0.5154(19)$, $\nu=1.33(7)$ and $\gamma=2.43(4)$ with quality $S=1.16$.

\textit{Average size of the finite clusters.}
The last observable describes the average size of all finite (non-percolating) clusters that appear in one instance of the GG. Naturally, this quantity is also averaged over $2000$ instances.
The definition is \cite{Stauffer1992,Sur1976}:
\begin{equation}
    S_{\rm fin}(p) = \frac{\sum^\prime s^2\,n_s(p)}{\sum^\prime s\,n_s(p)},
    \label{eq:sfin}
\end{equation}
where $n_s(p)$ denotes the probability mass function of cluster sizes for as single instance of the GG. 
The sum $\sum^\prime$ runs over all clusters except the percolating ones.
It is expected that this quantity scales similar to the fluctuations of the order parameter.
Considering $\epsilon\in[-1,1.5]$ we obtain $p_c=0.5164(11)$, $\nu=1.33(6)$ and $\gamma=2.39(5)$ with quality $S=0.71$.

\subsection{Results for site percolation on planar GGs}
\label{sect:site}

The analysis of site percolation on planar GGs is analogous to bond percolation (Sect.\ \ref{sect:bond}).
For that reason we do not depict the data plots, but show our results for the percolation thresholds.
We find $p_c=0.6340(11)$ (percolation probability), $p_c=0.6342(19)$ (binder ratio), $p_c=0.6348(8)$ (order parameter), $p_c=0.6332(21)$ (fluctuations of the order parameter) and $p_c=0.6340(15)$ (average size of the finite clusters).
All obtained estimates of the critical exponents (not shown here) are in agreement with the known analytical values.
The most accurate estimate of the critical points is obtained by studying the order parameter. This is also the case in the bond percolation study. 
For that reason we list these estimates in Tab.\ \ref{tab:critp}. 

\section{Summary}
\label{sect:summary}

In this article we have performed numerical simulations to find the bond and site percolation thresholds for the Gabriel graph. Recently, it has already been proofed \cite{Bertin2002} that there are nontrivial thresholds, but accurate estimates of these thresholds have not existed (to the best of our knowledge). 
In particular, there was no proper finite-size scaling analysis.
Considering different observables, we find $p_c = 0.5167(6)$ (bond percolation) and $p_c = 0.6348(8)$ (site percolation) by means of finite-size scaling analyses. 
We also find estimates of the critical exponents $\nu$, $\beta$ and $\gamma$ that are in good agreement with the known values obtained from standard $2D$ percolation.
The considered Gabriel graph is a subgraph of the Delaunay triangulation whose percolation thresholds ($p_{c,{\rm bond}}=0.333069(2)$ and $p_{c,{\rm site}}=0.5$) are already well understood \cite{Becker2009,Bollobas2006}.
Also the percolation thresholds of the relative-neighborhood graph  ($p_{c,{\rm bond}}=0.771(2)$ and $p_{c,{\rm site}}=0.796(2)$) are known yet \cite{Melchert2013-2}.
Following the containment principle due to Fisher \cite{Fisher1961}, it should hold that $p^{\rm RNG}_c \geq p^{\rm GG}_c \geq p^{\rm DT}_c$ for both percolation problems, which has been confirmed.
The averaged degree of the Gabriel graph is $4$.
Just for comparison, the percolation thresholds for the $2D$ square lattice, which also exhibits a degree of $4$, are slightly smaller: $p_{c,{\rm bond}}=0.5$ and $p_{c,{\rm site}}=0.59274621(13)$ \cite{Newman2000}.

\section{Acknowledgments}
I am very grateful to Oliver Melchert for valuable discussions and comments.
Financial support was obtained via the
Lower Saxony research network ``Smart Nord'' which
acknowledges the support of the Lower Saxony Ministry of Science
and Culture through the ``Nieders\"achsisches Vorab'' grant program
(grant ZN 2764/ ZN 2896).

\bibliography{research.bib}

\end{document}